\documentclass[journal,10pt]{IEEEtran}

\usepackage{stfloats}
\usepackage{amsmath}
\usepackage{amssymb,latexsym}
\usepackage{graphicx}
\usepackage{color,cite,times}
\usepackage{epstopdf}
\usepackage{algorithmic}
\usepackage[ruled,linesnumbered] {algorithm2e}
\usepackage{booktabs}
\usepackage{bbm} 
\usepackage{url}
\usepackage{fancyhdr}
\usepackage{verbatim}
\usepackage{multirow}
\usepackage{makecell}
\usepackage{graphicx}
\usepackage{indentfirst}
\usepackage{cases}
\newcommand{\RNum}[1]{\uppercase\expandafter{\romannumeral #1\relax}}
\usepackage{colortbl} 
\usepackage{extarrows}
\usepackage{threeparttable}
\usepackage{diagbox}
\usepackage{pifont}
\usepackage{cancel}
\usepackage{framed}
\usepackage{overpic}
\usepackage{mathtools} 
\usepackage{hyperref}
\hypersetup{
	colorlinks=true,
	linkcolor=blue,
	citecolor=blue,
	urlcolor=black
}
\usepackage{cleveref}
\usepackage{amsthm,amsmath}

\usepackage{subfigure}

\usepackage[framemethod=tikz]{mdframed}
\usepackage{lipsum}
\usepackage[nolist]{acronym}
\begin{acronym}

\acro{AI}{artificial intelligence}
\acro{NFE}{neural function evaluation}
\acro{LLM}{large language model}
\acro{NN}{neural network}
\acro{DSM}{denoising score matching}
\acro{SNR}{signal-to-noise ratio}
\acro{JADCE}{joint activity detection and channel estimation}
\acro{CRB}{Cram\'er--Rao bound}
\acro{LoS}{line-of-sight}
\acro{NLoS}{non-LoS}
\acro{UPA}{uniform planar array}
\acro{ML}{maximum-likelihood}
\acro{RMSE}{root MSE}
\acro{JSCC}{joint source-channel coding}
\acro{ISAC}{Integrated Sensing and Communication}
\acro{FPGA}{field-programmable gate array}
\acro{GPU}{graphics processing unit}

\acro{MISO}{multiple-input single-output}
\acro{MIMO}{multiple-input multiple-output}
\acro{SISO}{single-input single-output}
\acro{OFDM}{orthogonal frequency division multiplexing}

\acro{CT}{computed tomography}
\acro{SAR}{synthetic aperture radar}
\acro{MRI}{magnetic resonance imaging}

\acro{w.r.t.}{with respect to}

\acro{AMP}{approximate message passing}
\acro{D-AMP}{denoising-based AMP}
\acro{D-Turbo-CS}{denoising-based Turbo-CS}
\acro{BM3D-prGAMP}{BM3D phase-retrieval GAMP}
\acro{LDAMP}{learned D-AMP}
\acro{VAMP}{vector AMP}
\acro{OAMP}{orthogonal AMP}
\acro{BP}{belief propagation}
\acro{CS}{compressive sensing}
\acro{TV}{total variation}

\acro{VAE}{variational autoencoder}
\acro{GAN}{generative adversarial network}
\acro{MAMP}{memory AMP}
\acro{GAMP}{generalized AMP}

\acro{EP}{expectation propagation}
\acro{EC}{expectation consistent}
\acro{MSE}{mean-squared error}
\acro{MMSE}{minimum mean-squared error}
\acro{NMSE}{normalized MSE}
\acro{LMMSE}{linear MMSE}
\acro{SVD}{singular value decomposition}
\acro{KL}{Kullback-Leibler}
\acro{KKT}{Karush-Kuhn-Tucker}
\acro{AWGN}{additive white Gaussian noise}
\acro{ROI}{right-orthogonally invariant}
\acro{DCT}{discrete cosine transform}
\acro{DFT}{discrete Fourier transform}
\acro{i.i.d.}{independent and identically distributed}
\acro{SE}{state evolution}

\acro{STMP}{score-based turbo message passing}
\acro{Q-STMP}{quantized STMP}
\acro{ADCs}{analog-to-digital converters}
\acro{SDE}{stochastic differencial equation}
\acro{VE}{variance exploding}
\acro{GLMs}{generalized linear models}
\acro{SE}{state evolution}
\acro{TMP}{turbo message passing}

\acro{DnCNN}{denoising convolutional neural network}
\acro{E2E}{end-to-end}

\acro{PDF}{probability density function}
\acro{CDF}{cumulative distribution function}

\acro{MAP}{maximum \textit{a posteriori}}
\acro{FISTA}{fast iterative shrinkage-thresholding algorithm}
\acro{ADMM}{alternating direction method of multipliers}
\acro{RED}{regularization by denoising}
\acro{HQS}{half-quadratic splitting}
\acro{GTurbo-SR}{generalized turbo signal recovery}
\acro{PnP}{plug-and-play}

\acro{DPS}{diffusion posterior sampling}
\acro{DiffPIR}{diffusion models for PnP image restoration}
\acro{DDRM}{denoising diffusion restoration models}
\acro{DMPS}{diffusion model based posterior sampling}
\acro{MCG}{manifold constrained gradient}
\acro{Score-ALD}{score-based annealed Langevin dynamics}
\acro{ILVR}{iterative latent variable refinement}
\acro{DDIM}{denoising diffusion implicit models}
\acro{ODE}{ordinary differential equation}

\acro{MPDQ}{message-passing de-quantization}
\acro{QSP}{quantized subspace pursuit}
\acro{QIHT}{quantized iterative hard thresholding}
\acro{QCoSaMP}{quantized compressive sampling matching pursuit}

\acro{CNN}{convolutional neural network}

\acro{RMP}{reverse mean propagation}
\end{acronym}

\newcommand{\bsm}{\boldsymbol}

\definecolor{gray}{RGB}{192,192,192}

\begin{document}

\title{Wireless Intelligence Needs a Cerebellum: Score-Based Foundation Models Toward Real-Time Physical-Layer Inference}

\author{Chang Cai,
	Boyu Teng,
	Xiaojun Yuan, \IEEEmembership{Fellow, IEEE,} 
	and Ying-Jun Angela Zhang,  \IEEEmembership{Fellow, IEEE}  
		\thanks{
			Chang Cai is with the Department of Electrical and Computer Engineering, The University of Hong Kong, Hong Kong SAR (e-mail: changcai@hku.hk).
			
			Boyu Teng and Xiaojun Yuan are with the National Key Laboratory of Wireless Communications,
			University of Electronic Science and Technology of China, Chengdu 611731, China (e-mail: byteng@std.uestc.edu.cn; xjyuan@uestc.edu.cn).
			
			Ying-Jun Angela Zhang is with the Department of Information Engineering, The Chinese University of Hong Kong, Hong Kong SAR (e-mail: yjzhang@ie.cuhk.edu.hk).
			
		}
	}
\maketitle
\begin{abstract}
	Wireless intelligence requires not only large foundation models for network-wide planning and decision-making, but also a compact ``cerebellum'' for fast and precise physical-layer inference.
	Unlike the computation-intensive architectures used at upper layers, the physical-layer cerebellum must operate within stringent microsecond-to-millisecond latency constraints.
	This article presents ScoreFM, a lightweight score-based foundation model designed for this role.
	ScoreFM learns reusable score functions that characterize the priors of wireless channels, source signals, and structured interference.
	During inference, these learned priors are embedded into task-specific message-passing algorithms as plug-and-play denoisers, allowing the same compact score networks to support diverse downstream tasks.
	This design combines the expressive power of score-based generative learning with the efficiency, interpretability, and modularity of model-based inference.
	Case studies on channel estimation, localization, and blind semantic communication demonstrate the flexibility and effectiveness of ScoreFM.
	Finally, we discuss future directions and open challenges toward realizing a practical wireless cerebellum.
\end{abstract}

\section{Introduction}
The growing vision of \ac{AI}-native wireless networks has stimulated increasing interest in wireless foundation models that can support a broad range of communication, sensing, and networking tasks~\cite{wentao2026foundation_model, liangle2026icm}.
Most existing foundation-model paradigms, however, inherit a scale-centric philosophy from language and vision, where increasing the number of parameters and the amount of training data is often regarded as the primary route toward stronger general-purpose intelligence~\cite{tianyue2026llm_phy_layer}.
Such a philosophy does not directly match the operational characteristics of the wireless physical layer.
As the last-mile interface between digital intelligence and the radio environment, the physical layer must process radio-domain information under stringent timing constraints.
Operations such as data detection, channel estimation, interference suppression, and beam adaptation are executed within tightly constrained microsecond-to-millisecond control loops.
In such regimes, computational efficiency becomes a central requirement.
It is therefore impractical to directly place an increasingly large model inside every physical-layer control loop. 

\begin{figure}[t]
	\centering
	\includegraphics[width=1\columnwidth]{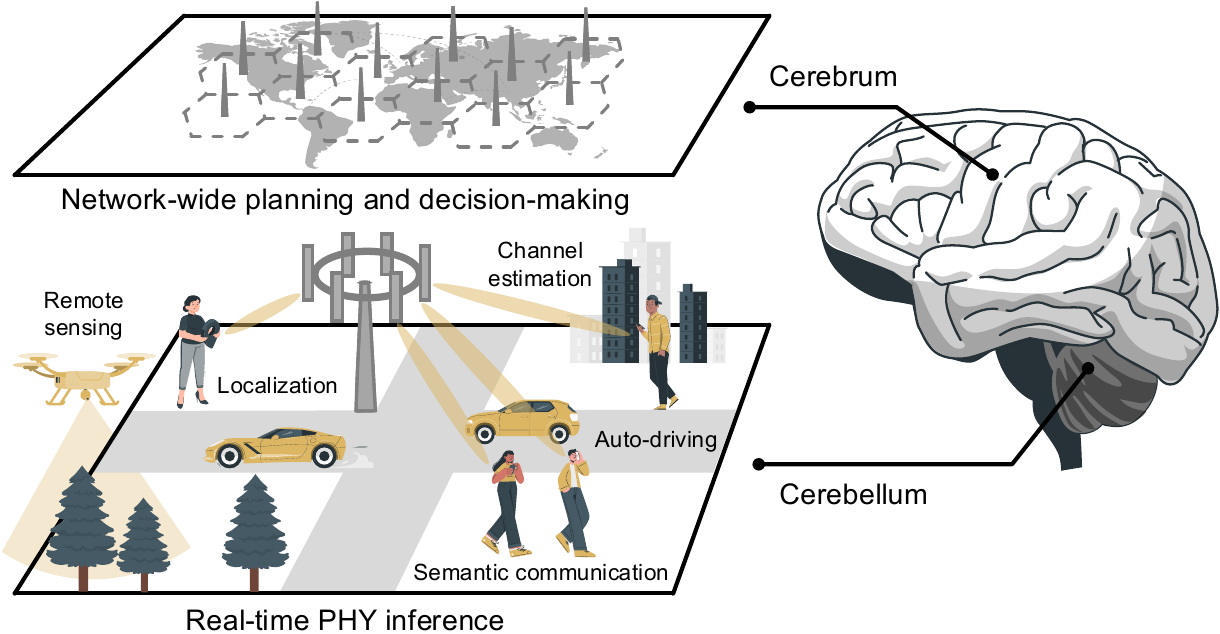}
	\caption{Conceptual illustration of wireless intelligence inspired by human brain.
		Large foundation models act as the wireless cerebrum, supporting network-wide planning and decision-making, while compact models serve as the wireless cerebellum, enabling real-time physical-layer inference.
	}
	\label{fig:brain}
\end{figure}

A more appropriate architecture may be found in the cooperation between the cerebrum and the cerebellum in biological intelligence~\cite{kandel2021principles}.
The cerebrum is responsible for high-level reasoning, semantic understanding, planning, and long-horizon decision-making.
In contrast, the cerebellum uses learned internal models to execute fast, precise, and repetitive sensorimotor control, without invoking complex deliberative reasoning for every action.
As illustrated in Fig.~\ref{fig:brain}, wireless intelligence can be organized in a similar hierarchical manner.
A large upper-layer model may interpret service requirements, coordinate network-wide resources, select operating strategies, and determine when adaptation is necessary.
Meanwhile, compact physical-layer models must execute fast inference and control close to the radio interface.
From this perspective, the central question is no longer how to construct a single model that is sufficiently large to solve every wireless task across different layers.
Instead, it is how to develop compact models that possess reusable knowledge, adapt to different physical-layer tasks, and operate within stringent real-time constraints.

Existing physical-layer foundation-model solutions do not fully satisfy these requirements.
One line of work adopts transformer-based \acp{LLM} as shared backbones, with task-specific output heads attached for different downstream tasks~\cite{liangle2026icm, tianyue2026llm_phy_layer}.
Although this paradigm facilitates representation reuse, it also inherits the complexity of large architectures and, in the case of autoregressive models, may further incur sequential decoding latency.
Moreover, adapting the shared backbone to a new physical-layer task typically requires additional fine-tuning.
Another line of work exploits generative models to learn the underlying distribution of wireless channels or signals independent of a specific inference task~\cite{langevin2023twc, mimoscore2023twc}.
Score-based diffusion models~\cite{song2021sde}, in particular, offer expressive priors that can be combined with different likelihood functions through Bayesian inference.
Their conventional use, however, relies on iterative reverse diffusion for posterior sampling, which may involve hundreds or even thousands of \acp{NFE}.
Such a computational burden is difficult to reconcile with fast physical-layer operation.

In this article, we present score-based foundation models, termed ScoreFM, for lightweight, task-adaptive, and real-time physical-layer inference.
Here, the foundation-model property refers to reusable statistical knowledge that supports multiple downstream tasks without retraining the underlying learned priors, rather than to a single monolithic backbone.
ScoreFM employs separate compact \acp{NN} to learn the score functions of physical variables, including wireless channels, source signals, and structured interference.
Each score function represents the gradient of the log-density of its corresponding prior distribution.
These learned priors are decoupled from task-dependent likelihoods and physical constraints, allowing the same score network to be combined with different observation models across tasks.
To enable low-latency operation, ScoreFM embeds the priors into model-based iterative algorithms, particularly message-passing frameworks~\cite{cai2025spawc}, where they serve as plug-and-play denoisers.
Rather than executing a full reverse-diffusion trajectory, each score network is evaluated only once in each message-passing iteration, substantially reducing the number of \acp{NFE}.
This design exploits the expressive power of score-based generative learning while retaining the efficiency, interpretability, and modularity of model-based inference.
The case studies on channel estimation, localization, and blind semantic communication illustrate this reusable-prior paradigm, while the subsequent discussion examines lightweight implementations and remaining challenges in generalization and scalability.

\section{ScoreFM Toward Real-Time Physical-Layer Inference}
This section details the construction of ScoreFM.
We first cast a broad class of physical-layer inference tasks into a unified Bayesian formulation, where task-dependent likelihoods characterize the observation process and reusable priors capture the statistical structures of channels, source signals, and interference.
We then show that the resulting probability factorization admits a message-passing implementation, in which learned score priors act as plug-and-play denoisers across diverse physical-layer inference tasks.

\subsection{Physical-Layer Inference Problem}
Consider a representative abstraction of a wireless system: $\bsm{Y} = \bsm{H} \bsm{X} + \bsm{Z} + \bsm{N}$, where $\bsm{Y}$ denotes the received signal, $\bsm{H}$ the wireless channel, $\bsm{X}$ the transmitted signal, $\bsm{Z}$ an interference component, and $\bsm{N}$ the \ac{AWGN}.
The transmitted signal $\bsm{X}$ is generated from an underlying source $\bsm{S}$ through an encoding process $\bsm{X} = f(\bsm{S})$, which may represent source coding, channel coding, modulation, or task-oriented semantic encoding.
This model captures a broad class of communication and sensing problems in which the received signal is determined by transmitted symbols, propagation channels, structured interference, and thermal noise. Its precise dimensions, variable interpretations, and factorization may vary across applications, and additional latent variables or coupling factors can be introduced when required.
For instance, $\bsm{X}$ may represent a single-user symbol sequence, a multiuser transmit vector, or spatial streams across multiple antennas, while $\bsm{H}$ captures the corresponding propagation effects in time, frequency, or space.
The formulation can also be extended to tensor observation models that arise in multi-dimensional signal processing problems.
This unified abstraction provides a convenient starting point for developing a general inference framework applicable to a wide range of physical-layer signal processing tasks.

Under this abstraction, different physical-layer tasks can be viewed as estimating different subsets of $\left\{\bsm{H}, \bsm{S}, \bsm{X}, \bsm{Z}\right\}$ from $\bsm{Y}$, depending on which variables are known, unknown, or treated as nuisance components.
Examples include:
\begin{itemize}
	\item \textbf{Channel estimation:} the transmitted signal $\bsm{X}$ consists of known pilot symbols, and the goal is to estimate the channel $\bsm{H}$ from the observation $\bsm{Y}$.
	
	\item \textbf{Data detection:} the channel $\bsm{H}$ is assumed known (or previously estimated), and the goal is to recover the transmitted symbols $\bsm{X}$ from $\bsm{Y}$.
	
	\item \textbf{Interference suppression:} the interference component $\bsm{Z}$ is treated as an unknown structured signal that needs to be estimated or mitigated.
	
	\item \textbf{Wireless sensing and localization:} the unknown variables $\bsm{H}$ and $\bsm{Z}$ encode physical parameters such as angles, delays, and target reflections, which can be inferred from $\bsm{Y}$.
	
	\item \textbf{Semantic communication:}
	the encoder $f(\cdot)$ maps the source signal $\bsm{S}$ into a transmitted representation $\bsm{X}$ that preserves task-relevant semantic information.
	The receiver aims to infer $\bsm{S}$, or more commonly a task-relevant function of $\bsm{S}$, from $\bsm{Y}$.
\end{itemize}
More challenging tasks arise when multiple unknown variables must be inferred jointly.
Representative examples include blind detection, where the channel and transmitted signal are estimated simultaneously, and \ac{JADCE} in massive connectivity systems, where the activity states of sporadically accessing users and their associated channel coefficients are recovered together.

From a Bayesian perspective, these problems can be formulated as inferring the unknown variables from the observation $\bsm{Y}$.
This corresponds to evaluating the joint posterior distribution $p(\bsm{H}, \bsm{S}, \bsm{X}, \bsm{Z}|\bsm{Y})$, or its marginal distributions for the variables of interest.
By Bayes' rule, the joint posterior can be factorized as $p(\bsm{H}, \bsm{S}, \bsm{X}, \bsm{Z}|\bsm{Y}) \propto p(\bsm{Y}| \bsm{H}, \bsm{X}, \bsm{Z}) p(\bsm{H}) p (\bsm{S})p(\bsm{X}|\bsm{S}) p(\bsm{Z})$.
Here, $p(\bsm{H})$, $p(\bsm{S})$, and $p(\bsm{Z})$ represent the prior distributions of the channel, source signal, and interference, respectively; $p(\bsm{Y}|\bsm{H},\bsm{X},\bsm{Z})$ is the likelihood determined by the system model; and $p(\bsm{X}|\bsm{S})$ describes the encoding process.
This factorization serves as a baseline abstraction for exposition.
More general dependencies can be represented by introducing additional coupling factors without changing the basic ScoreFM principle.

In practice, the most challenging part of the inference problem is rarely the observation model.
In many wireless systems, the likelihood function is well understood and often admits convenient analytical forms, such as Gaussian models arising from linear or locally linear observation relationships.
The main challenge instead lies in modeling the prior distributions of the unknown variables.
In realistic wireless environments, these priors capture complex phenomena such as structured source signals, spatially and temporally correlated propagation channels, and aggregated inter-cell interference.
As a result, the underlying distributions are typically high-dimensional, strongly correlated, and highly non-Gaussian, making them extremely difficult to characterize analytically.

These modeling challenges motivate data-driven solutions.
Conventional learning-based methods are typically designed for a single task under a specific system configuration.
Adapting them to different propagation environments, \acp{SNR}, antenna configurations, or measurement models typically requires separate training, which limits transferability to unseen conditions.
Wireless foundation models instead seek to overcome this limitation by learning reusable representations or statistical knowledge from large-scale wireless data, thereby providing a common backbone for a broad range of downstream tasks.
Existing approaches generally rely either on \ac{LLM}-based backbones with task-specific adaptation~\cite{liangle2026icm,tianyue2026llm_phy_layer}, or on generative priors, particularly score-based diffusion models, combined with task-dependent likelihoods~\cite{wentao2026foundation_model, langevin2023twc,mimoscore2023twc}.
The former inherits the computational complexity of large architectures, while the latter typically requires many iterative sampling steps.
Consequently, both paradigms face difficulty meeting the stringent latency requirements of the physical layer.

\subsection{ScoreFM Framework}
Motivated by the above, we propose ScoreFM, which integrates lightweight score networks into model-based message-passing algorithms for efficient and task-adaptive physical-layer inference.
The score networks capture task-independent statistical priors of source signals, wireless channels, and interference. 
These reusable score-based priors constitute the foundation knowledge in ScoreFM and are combined with task-dependent observation models during inference.
Message passing exploits the posterior factorization to coordinate information exchange among the observation, encoding, and score-based prior factors, thereby decomposing a complex inference problem into tractable estimation modules.
This modular design allows the learned priors to be reused across diverse downstream tasks and system configurations, without relying on large \ac{E2E} architectures or lengthy reverse-diffusion trajectories.
We present the learning and inference procedures of ScoreFM as follows.

\begin{figure}[t]
	\centering
	\includegraphics[width=1\columnwidth]{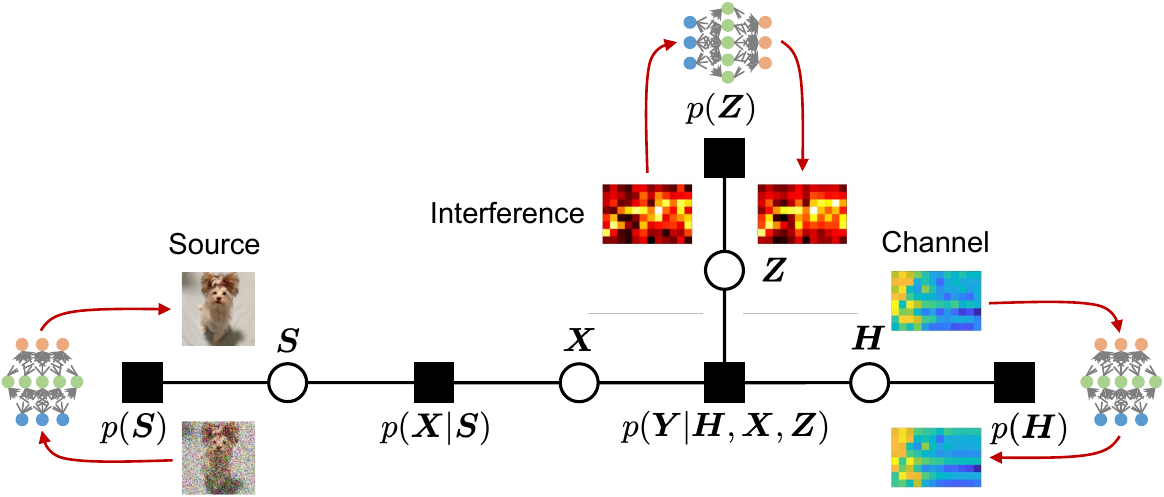}
	\caption{Factor-graph representation of the unified Bayesian inference problem, where circular nodes represent variable nodes corresponding to the unknown variables, while square nodes represent factor nodes associated with probabilistic factors such as the likelihood and prior distributions.
		An edge connects a variable node and a factor node if the variable participates in that factor.}
	\label{fig:factor_graph_framework}
\end{figure}

\subsubsection{Learning} 
ScoreFM learns each reusable prior independently of any specific downstream task or observation model.
Given samples of source signals, wireless channels, or interference components, a compact \ac{NN} is trained via \ac{DSM}~\cite{song2021sde} to estimate the score function of the corresponding Gaussian-perturbed distribution over a range of noise levels.
Specifically, clean samples are corrupted by additive Gaussian noise, and the network is conditioned on the noise level to predict the gradient of the log-density of the perturbed data.
This training process requires only samples from the underlying prior and does not rely on paired observations or task-specific labels.
In practice, score networks commonly use a lightweight U-Net architecture~\cite{song2021sde} composed of stacked convolutional blocks in an encoder-decoder structure.
Skip connections fuse multiscale features across corresponding stages, helping capture both local structures and long-range correlations in high-dimensional wireless data.
Once trained, the score network is frozen and can be reused across different inference tasks and observation configurations, provided that the underlying prior distribution remains approximately unchanged. This reuse primarily concerns task generalization; adaptation across substantially different physical domains is discussed in Section~\ref{subsec:open_challenges}.

\subsubsection{Inference}
During inference, ScoreFM combines the learned priors with task-specific observation models through message passing.
The factorization of the joint posterior naturally gives rise to the factor graph shown in Fig.~\ref{fig:factor_graph_framework}.
Message-passing algorithms operate on this factor graph by iteratively exchanging probabilistic messages between variable and factor nodes, thereby producing approximate marginal posteriors for the unknown variables.
The framework is readily adapted to different inference tasks by retaining only the relevant branches of the graph.
For example, when the channel is known, the branch associated with $p(\bsm{H})$ can simply be removed, while the remaining branches estimate the source signal and suppress the interference.
Likewise, variables with tractable priors can be modeled explicitly, whereas only analytically intractable priors are replaced by learned score networks.
This flexibility enables the integration of model-driven and data-driven components within a unified Bayesian inference framework.

A key advantage of ScoreFM lies in its modular inference architecture, in which the factor graph is organized into modules responsible for different components of the probabilistic model.
The central idea is that message passing converts the original global inference problem into a sequence of simpler denoising steps.
At each learned-prior factor, information collected from the rest of the graph can be viewed as an effective noisy estimate of the corresponding variable.
The score network then refines this estimate using the learned prior at the appropriate noise level.
In this way, ScoreFM uses the score network directly as a denoiser inside the inference loop, requiring one \ac{NFE} per prior module per iteration rather than repeated evaluations along a full reverse-diffusion trajectory.
Other modules handle the observation model and additional probabilistic constraints through analytical or model-based updates.
Together, these modules decompose the global inference problem into a collection of tractable local subproblems.
They iteratively exchange messages along the factor graph, progressively refining the approximate marginal beliefs of the unknown variables.
For the ScoreFM instances considered in the following case studies, the underlying message-passing algorithms converge within a small number of iterations.
ScoreFM therefore requires only a small number of score-network evaluations per prior module.

\begin{figure*}[t]
	\centering
	\includegraphics[width=1.6\columnwidth]{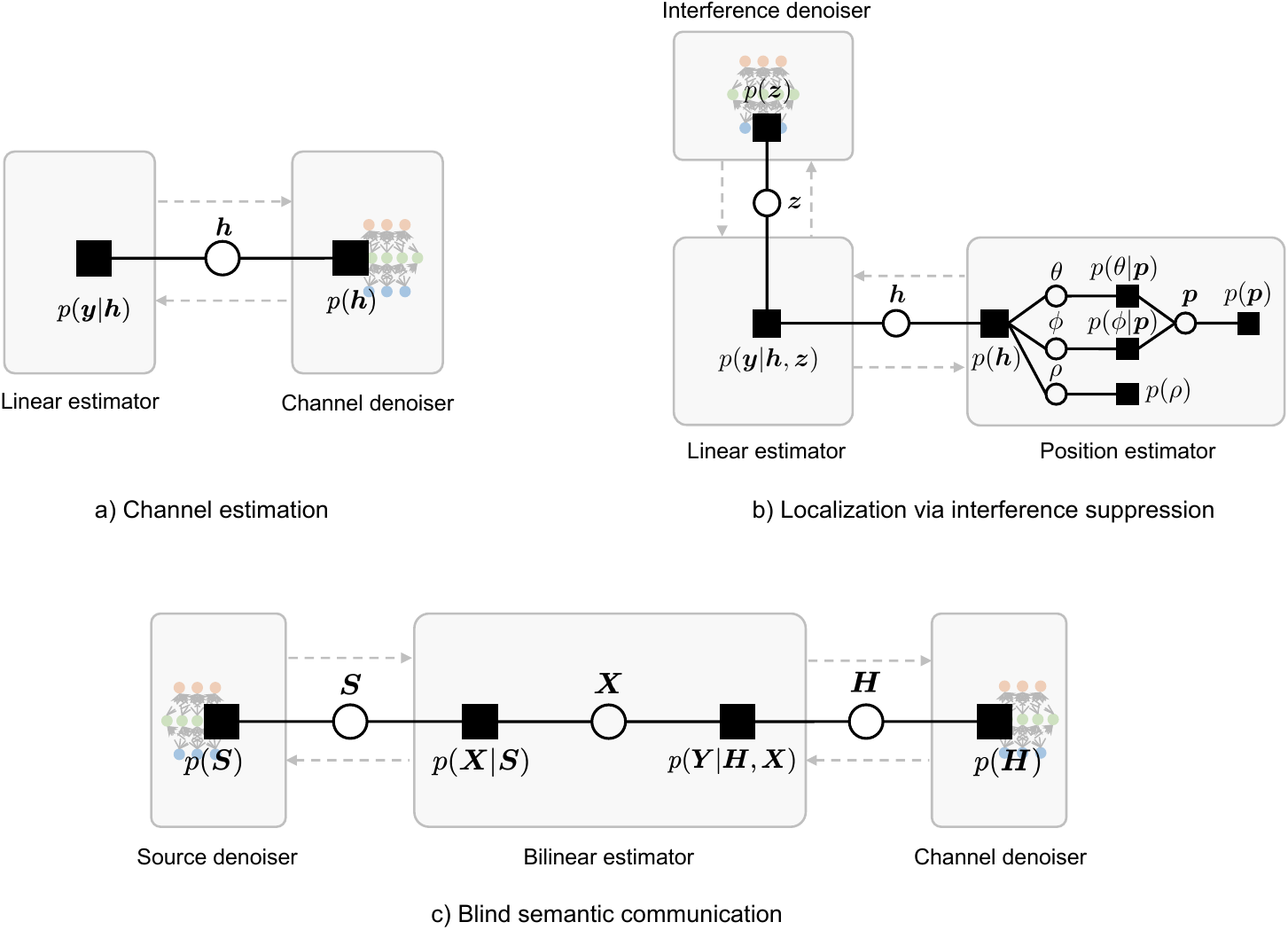}
	\caption{Instantiations of ScoreFM for three physical-layer inference tasks. Learned score priors are reused as denoisers, while task-dependent observation and physical models are handled by analytical or model-based estimators.}
	\label{fig:case_studies}
\end{figure*}

\section{Case Studies}

\subsection{Channel Estimation}

We first consider channel estimation in a \ac{MISO} downlink system.
With no interference and all transmitted pilot symbols known at the receiver, the factor graph reduces to a linear observation factor connected to the channel prior, as illustrated in Fig.~\ref{fig:case_studies}(a).
Despite this relatively simple inference structure, accurate channel estimation remains challenging because the number of available pilot observations is often substantially smaller than the channel dimension.
The resulting underdetermined problem is therefore highly dependent on the accuracy and expressiveness of the adopted channel prior.

To address this challenge, our framework interprets the channel prior as a denoiser operating on an effective Gaussian observation produced by message passing.
This yields a modular architecture comprising a linear estimator and a channel denoiser~\cite{cai2025spawc}.
The linear estimator incorporates the pilot observations, while the denoiser exploits prior knowledge of the channel structure.
By representing this prior with a learned score-based generative model, the denoiser can capture complex, high-dimensional channel statistics beyond conventional parametric models.
The two components iteratively exchange extrinsic information in a turbo manner, progressively refining the channel estimate.

\begin{figure}[t]
	\centering
	\includegraphics[width=1\columnwidth]{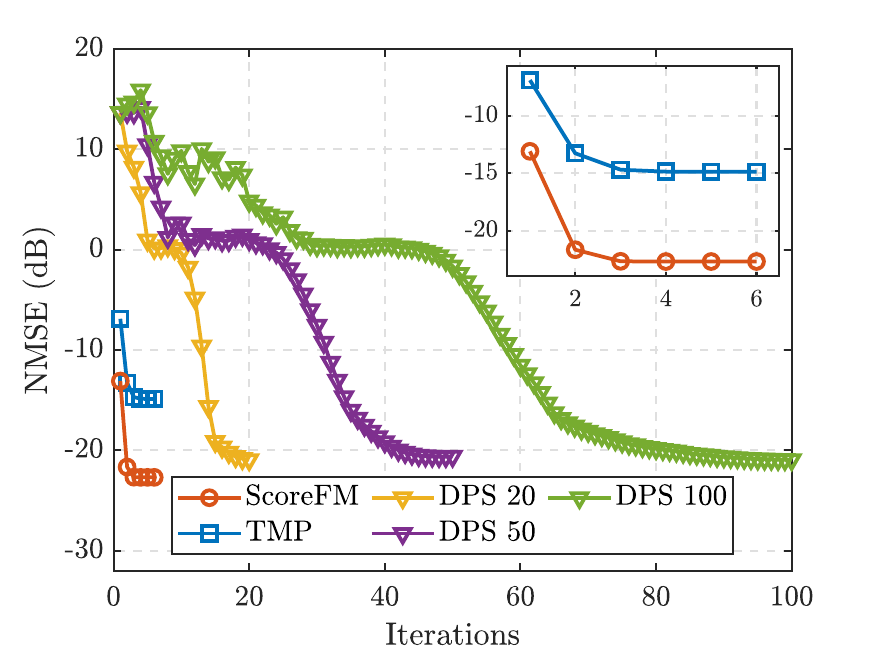}
	\caption{\Ac{NMSE} convergence of channel estimation in a \Ac{MISO} downlink system with $256$ transmit antennas.
		The pilot ratio is $0.6$, and the \ac{SNR} is $10$ dB.
		\Ac{DPS} baselines use 20, 50, and 100 reverse-diffusion steps.
		}
	\label{fig:channel_estimation}
\end{figure}

We evaluate ScoreFM using channels generated from a mixture of the 3GPP CDL-A through CDL-E models, which capture diverse propagation conditions ranging from sparse \ac{LoS} scenarios to rich \ac{NLoS} multipath environments.
We instantiate the score prior using the NCSN++ architecture~\cite{song2021sde}, which has been widely adopted in score-based generative modeling.
As shown in Fig.~\ref{fig:channel_estimation}, ScoreFM achieves significantly faster convergence than \acf{DPS} baselines. 
In particular, the proposed method reaches a low-\ac{NMSE} operating point with only 2--3 score-network evaluations, whereas \ac{DPS} requires many more evaluations to approach comparable accuracy.
This is because \ac{DPS} enforces measurement consistency through gradient guidance along the diffusion trajectory, which typically requires a much longer sampling chain.
Fig.~\ref{fig:channel_estimation} also shows that ScoreFM substantially outperforms \ac{TMP} with Bernoulli-Gaussian sparsity priors in the angular domain.
This demonstrates that the learned score prior captures channel structures beyond conventional sparse parametric models.

\begin{table}[t]
	\caption{\Ac{NMSE} of channel estimation achieved by score networks of different sizes. The gap column reports the \ac{NMSE} difference relative to the full NCSN++ model.}
	\label{table:nmse_comparison}
	\centering
	\begin{tabular}{cccc}
		\toprule
		Model & Size (M) & NMSE (dB) & Gap (dB) \\
		\midrule
		\multirow{5}{*}{Lightweight}& 0.306 & $-20.793 \pm 0.944$ & $+1.587 \pm 0.389$ \\
		& 0.368 & $-20.837 \pm 0.871$ & $+1.544 \pm 0.345$ \\
		& 0.587 & $-20.878 \pm 0.835$ & $+1.503 \pm 0.363$ \\
		& 0.845 & $-21.147 \pm 0.844$ & $+1.234 \pm 0.379$ \\
		& 1.22 & $-21.309 \pm 0.855$ & $+1.072 \pm 0.338$ \\
		\midrule
		NCSN++ & 42.6 & $-22.381 \pm 0.977$ & -- \\
		\bottomrule
	\end{tabular}
\end{table}

To support real-time inference, we further reduce the size of the score network and evaluate the resulting impact on estimation accuracy.
Existing architectures such as NCSN++ were originally developed for high-resolution computer vision tasks and may be overparameterized for the channel representations considered here.
These representations exhibit strong domain-specific structure and can often be modeled effectively with substantially smaller networks.
We therefore construct lightweight score networks by reducing the base channel width and the number of residual blocks. 
As shown in Table~\ref{table:nmse_comparison}, the full NCSN++ model contains more than 42 million parameters, whereas compact variants with only 0.3--1.2 million parameters achieve competitive \ac{NMSE}, incurring only a modest performance loss despite a substantial reduction in model size.
These model sizes are also comparable to those used in existing real-time physical-layer implementations.
For example, TensorRT-optimized channel-estimation networks with approximately 0.1--0.9 million parameters have achieved inference latencies of 0.17--0.32~ms on a high-end \ac{GPU}~\cite{camelo2026helena}.
With customized \ac{FPGA} acceleration, neural receiver models containing 0.95--1.89 million parameters and 1.16--2.10 million multiply--accumulate operations have reported processing latencies of 10.85--39.44~$\mu$s~\cite{luo2025end}.
Although actual latency depends not only on parameter count but also on operation count, memory access, numerical precision, network architecture, and hardware mapping, these results provide useful evidence that networks at the scale considered in Table~\ref{table:nmse_comparison} are compatible with sub-millisecond physical-layer processing.

We further note that the same score-based channel prior can be reused for different inference tasks, such as \ac{JADCE} in grant-free massive random access.
In this setting, only a small and unknown subset of devices is active, and the receiver must jointly identify these devices and estimate their channels from limited pilot observations.
Compared with conventional channel estimation, the factor graph is augmented by introducing an activity variable for each device, together with a coupling factor that links the activity state to the corresponding channel.
This augmentation preserves the modular architecture of ScoreFM while extending it to jointly infer device activity and channel states.
Specifically, the linear estimator extracts channel observations from the received pilots, the score-based channel denoiser refines these observations through \ac{MMSE} denoising under the learned channel prior, and the activity estimator updates the posterior probability of each device being active.
Further details and numerical results are provided in~\cite{jadce2026jsac}.

\subsection{Interference-Aware Localization}
We next demonstrate how can ScoreFM combine explicit \ac{LoS} channel modeling with a learned interference prior.
We consider uplink localization~\cite{tby2026wcl} from a received array signal comprising the \ac{LoS} channel of the serving user, aggregated \ac{NLoS} multipath and co-channel interference, and thermal noise.
The \ac{LoS} component is explicitly parameterized by its complex gain and array steering vector, with the angular parameters linked to the user position through known geometric relationships.
By contrast, the aggregated interference is highly non-Gaussian and spatially correlated.
Approximating it as spatially white Gaussian noise may bias the \ac{LoS} parameter estimates.
The corresponding factor graph therefore retains an explicit geometry-aware model for the \ac{LoS} component while representing the interference through a learned score prior.
As illustrated in Fig.~\ref{fig:case_studies}(b), the resulting architecture comprises a linear estimator that separates the \ac{LoS} and interference components, a position estimator that extracts the \ac{LoS} angles and maps them to the user position, and a score-based interference denoiser.
These components iteratively exchange extrinsic information to suppress structured interference and progressively refine the position estimate.

\begin{figure}[t]
	\centering
	\includegraphics[width=1\columnwidth]{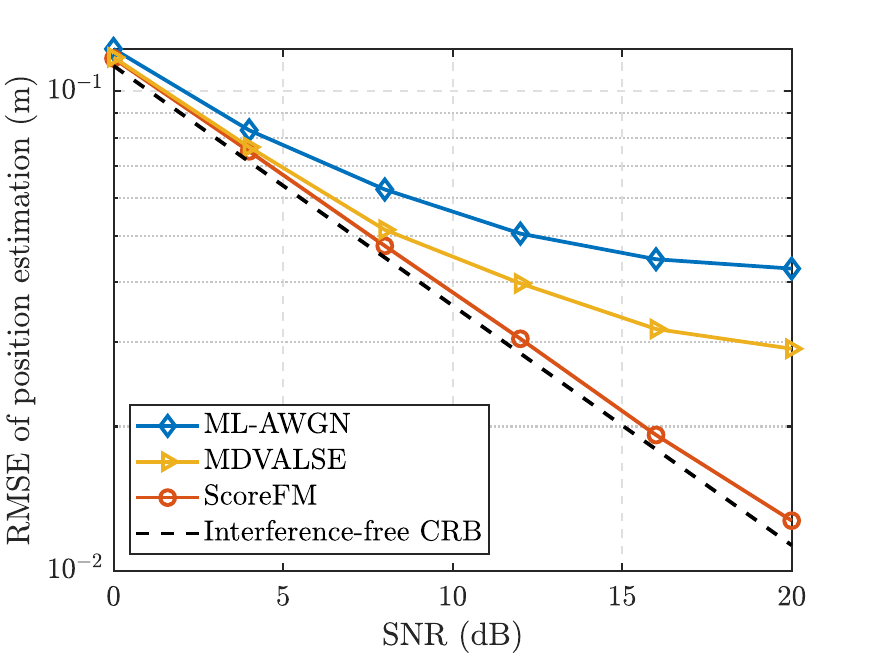}
	\caption{Localization \ac{RMSE} versus \ac{SNR} in a 28-GHz uplink system with a $64\times16$ \ac{UPA} and three interfering users.}
	\label{fig:localization}
\end{figure}

We evaluate ScoreFM for localization in a 28-GHz uplink system with a $64\times 16$ \acf{UPA}.
The interference comprises the \ac{NLoS} paths of the serving user and the signals from three interfering users.
These components are generated according to the 3GPP CDL-C model, with a total of 480 propagation paths.
As shown in Fig.~\ref{fig:localization}, ScoreFM consistently achieves a lower localization \acf{RMSE} than both ML-AWGN, which treats the interference as spatially white Gaussian noise, and MDVALSE, a variational line-spectrum estimation baseline.
As the \ac{SNR} increases, both baselines exhibit interference-limited error floors, whereas ScoreFM continues to improve and closely approaches the interference-free \ac{CRB}.
These results show that learning the structured interference prior effectively mitigates interference-induced bias, allowing the explicitly modeled \ac{LoS} geometry to be fully exploited for localization.

\subsection{Blind Semantic Communication}
This case study illustrates how multiple score networks can cooperate within ScoreFM to solve a joint inference problem. 
We consider blind semantic communication over a block-fading \ac{MIMO} channel, where the receiver must recover both the source data and the unknown channel without pilot assistance.
In conventional pilot-based systems, the pilot overhead grows with the number of transmit antennas and may consume a substantial fraction of the limited channel coherence interval.
Blind detection avoids dedicated pilots by jointly estimating the channel and transmitted signal, but the resulting bilinear factorization is generally non-unique.
Existing approaches therefore impose simplified sparsity assumptions and often require additional reference symbols to resolve permutation and phase ambiguities.
Semantic communication provides richer structural information for disambiguating the bilinear factorization: the source data exhibit strong correlations, while the \ac{JSCC} encoder maps them into a constrained set of codewords.
Together with the structured channel prior, these properties make completely pilot-free joint channel-and-source recovery possible~\cite{jiang2025blind}.

As illustrated in Fig.~\ref{fig:case_studies}(c), the factor graph for blind semantic communication is organized into three components comprising a source denoiser, a bilinear estimator, and a channel denoiser.
The bilinear estimator incorporates the known \ac{JSCC} encoding relationship and the \ac{MIMO} observation model, thereby coupling the source and channel estimates through the received signal.
Two independently trained score networks are employed to characterize the source and channel priors.
During inference, the bilinear estimator enforces consistency with the encoder and received observation, while the two score-based denoisers refine the source and channel estimates using their respective learned priors.
In this way, multiple score networks can operate jointly within the same factor graph.

\section{Future Directions and Open Challenges}
This section outlines promising directions for extending the framework and discusses unresolved challenges in implementation, generalization, and scalability.
\subsection{Future Directions}
\subsubsection{ScoreFM for Transceiver Co-Design}
The current ScoreFM framework primarily focuses on receiver-side inference under a fixed observation model.
An important future direction is to incorporate transmitter-side design, such as pilot construction, waveform optimization, resource allocation, and beam training.
Extending ScoreFM to the transmitter would allow learned score priors to support not only inference, but also the optimization of transmission and sensing strategies under task-specific objectives~\cite{linchen2025jcin}.
For instance, pilots, beams, and waveforms could be jointly designed with the inference procedure to directly improve downstream communication and sensing performance.
Moreover, posterior uncertainty provided by ScoreFM could further support adaptive designs, allowing subsequent transmission actions to focus on the most uncertain or decision-relevant dimensions.
This development would transform ScoreFM from a passive inference framework to an active transceiver co-design paradigm that jointly determines how physical-layer information is acquired, processed, and exploited.

\subsubsection{ScoreFM for Environment-Aware \ac{ISAC}}
Future \ac{ISAC} systems may benefit from explicitly modeling the physical environment rather than treating it only through its induced channel responses.
ScoreFM can learn score-based priors that characterize environmental structures, including building geometry, terrain morphology, vegetation cover, and atmospheric conditions.
Such priors can provide spatial context for localization, target detection and tracking, environmental mapping, and multipath-assisted sensing, particularly when direct observations are incomplete or severely occluded.
For example, learned geometric regularities can help infer unobserved regions of a radio map, associate \ac{NLoS} paths with likely reflecting surfaces, and constrain target trajectories according to the surrounding scene.
As sensing measurements accumulate over time, the environmental prior can also be progressively updated, enabling ScoreFM to maintain an evolving statistical representation of the physical world. 

\subsubsection{Temporally Adaptive ScoreFM}
The priors considered in the current framework mainly characterize static marginal distributions of channels, source signals, and interference.
Practical wireless environments, however, evolve over time because of user mobility, blockage, traffic variations, and changes in the surrounding geometry.
A promising direction is therefore to develop ScoreFM architectures that capture temporal dependencies by conditioning their priors on historical observations and contextual information, such as user location, velocity, carrier frequency, and past channel or beam estimates.
Such conditional priors could support channel prediction, beam tracking, and the suppression of temporally correlated interference.
Historical posterior estimates could also be used to initialize ScoreFM inference at subsequent time instants, thereby reducing pilot overhead and inference latency.

\subsubsection{Cross-Layer Cerebrum--Cerebellum Collaboration}
The cerebrum--cerebellum analogy suggests a hierarchical wireless-intelligence architecture in which large upper-layer models and compact physical-layer models operate at different timescales and assume complementary roles.
An upper-layer foundation model may interpret service requirements, coordinate network-wide resources, and determine the appropriate physical-layer operating strategy, while ScoreFM executes the corresponding inference and control tasks within stringent latency constraints.
A bidirectional interface is therefore needed between the two levels.
The upper layer can specify task objectives, allocate time-frequency and spatial resources, select the relevant physical-layer functions, and configure latency--reliability tradeoffs.
In return, ScoreFM can provide channel and interference estimates, posterior uncertainties, decoding outcomes, and other link-level performance indicators.
Such feedback allows the upper layer to revise its plans, reschedule users, reallocate resources, or trigger additional sensing and training when necessary.
This closed-loop collaboration provides a concrete pathway toward hierarchical wireless intelligence, rather than relying on a single monolithic model to address all tasks and timescales.

\subsection{Open Challenges}\label{subsec:open_challenges}
\subsubsection{Identifiability and Fundamental Limits of Joint Inference}
Although the modular factor-graph architecture allows multiple learned priors to be combined, it does not guarantee that all unknown variables can be recovered jointly.
Simultaneous estimation of the source, channel, and interference may suffer from scaling, phase, permutation, and decomposition ambiguities, since different combinations of these variables can produce similar received signals.
It remains unclear when the structural information captured by their respective priors is sufficient to resolve such ambiguities.
Important open questions include the identifiability conditions, minimum observation requirements, and achievable estimation accuracy under different prior strengths, noise levels, and system dimensions.
Establishing sample-complexity results, phase-transition characterizations, and Bayesian performance bounds would help distinguish fundamentally infeasible problems from those that are merely difficult to solve algorithmically.
The effects of score-model errors and prior mismatch on these limits also remain largely unexplored.

\subsubsection{Validity of Gaussian Denoising Interpretation}
ScoreFM relies on interpreting the messages entering each prior factor as effective Gaussian observations, which enables score-based MMSE denoising through Tweedie's formula.
This interpretation is well established for \ac{AMP} and certain \ac{EP}-based algorithms under suitable assumptions on the system dimensions and measurement matrices.
It may not remain valid for general factor graphs, particularly those involving nonlinear or bilinear factors, short loops, finite-dimensional systems, and strongly coupled variables.
In such cases, the effective errors may be colored, heteroscedastic, non-Gaussian, or correlated across iterations. 
Applying an isotropic Gaussian denoiser may then produce biased estimates, inaccurate uncertainty measures, or unstable message updates.
Further research is needed to characterize when the denoising interpretation remains accurate, develop practical diagnostics for detecting its breakdown, and extend ScoreFM to covariance-aware or non-Gaussian messages.

\subsubsection{Generalization Across Physical Domains}
As noted above, ScoreFM primarily supports task generalization by decoupling reusable priors from task-dependent likelihoods.
A score prior can therefore be reused across inference tasks, pilot patterns, or noise levels when the underlying data distribution remains approximately unchanged.
Generalization across physical domains is substantially more challenging.
Transitions between far-field and near-field propagation, different carrier frequencies, and varying antenna numbers, geometries, spacings, and apertures may change both the channel distribution and its representation.
Addressing these variations may require geometry-aware score networks conditioned on antenna coordinates and system parameters, together with architectures capable of handling variable-dimensional inputs.
Moreover, lightweight mechanisms such as context conditioning, low-rank adaptation, and mixtures of specialized score priors may be used to adapt a pre-trained model to previously unseen propagation conditions.

\subsubsection{Scalability of Score Priors}
Scaling ScoreFM to extremely large antenna arrays, wideband channels, and long source sequences remains a major challenge.
Although message passing reduces the number of network evaluations, applying score models directly in the original signal space may still incur prohibitive memory and latency.
Latent diffusion models~\cite{Rombach_2022_CVPR} provide a promising alternative, but it remains unclear how to construct latent representations that are sufficiently compact while preserving the information required for downstream physical-layer inference.
Moreover, the observation model and message updates are typically defined in the original signal domain, making principled inference and uncertainty propagation in the latent space difficult.
Patch-wise and factorized score models may offer complementary solutions, but their effects on estimation accuracy and statistical consistency are not yet well understood.
Resolving these issues is essential for deploying ScoreFM in large-scale real-time wireless systems.

\section{Concluding Remarks}
This article introduced ScoreFM as a compact foundation-model paradigm toward real-time physical-layer inference.
It learns reusable score-based priors for wireless channels, source signals, and structured interference.
By combining these learned priors with task-specific observation models through message passing,
ScoreFM achieves fast convergence and requires only one network evaluation per prior component in each iteration.
More broadly, ScoreFM suggests that physical-layer foundation models can be built from compact learned priors and structured inference algorithms rather than large \ac{E2E} architectures.
Further advances in implementation, generalization, and scalability will be essential for realizing a practical wireless cerebellum.
\bibliographystyle{IEEEtran}
\bibliography{IEEEabrv,mybib}

\end{document}